\documentclass[3p,onecolumn,preprint]{elsarticle}          % onecolumn

\usepackage[english]{babel}
\usepackage[utf8]{inputenc}
\usepackage[T1]{fontenc}
\usepackage{siunitx}
\usepackage{graphicx}
\usepackage{mathptmx}      % use Times fonts if available on your TeX system
\usepackage{amsmath}
\usepackage{amssymb}
\usepackage[colorlinks,citecolor=blue,urlcolor=blue,linkcolor=blue]{hyperref}
\usepackage{placeins}
\biboptions{numbers,sort&compress,square}
\usepackage[colorinlistoftodos,textsize=tiny]{todonotes}
\usepackage{soul}
\usepackage{multirow}
\usepackage{comment}
\newcommand{\amuhlo}{$a_{\mu}^\text{HLO}$}
\newcommand{\dalphahad}{$\Delta\alpha_{had}(t)$}
\newcommand{\amui}{$a_{\mu}^\text{HLO (Int)}$}
\newcommand{\amude}{$a_{\mu}^\text{HLO (Der)}$}
\newcommand{\amuip}{$a_{\mu}^\text{HLO, INPUT} = 695.1 \,\, (10^{-10})$}
\newcommand{\amuipfj}{$a_{\mu}^\text{HLO, INPUT} = 690.3 \,\, (10^{-10})$}
\newcommand{\amuipknt}{$a_{\mu}^\text{HLO, INPUT} = 693.0 \,\, (10^{-10})$}

\newcommand{\amud}{$a_\mu^{\text{HLO (I)}}$  ($10^{-10}$)}
\newcommand{\amup}{$a_\mu^{\text{HLO (II)}}$  ($10^{-10}$)}
\newcommand{\amuq}{$a_\mu^{\text{HLO (III)}}$  ($10^{-10}$)}
\newcommand{\amur}{$a_\mu^{\text{HLO (IV)}}$  ($10^{-10}$)}

\newcommand{\sa}{$s_0$ values}
\newcommand{\cA}{{\cal A}}
\newcommand{\cB}{{\cal B}}
\newcommand{\cS}{{\cal S}}
\newcommand{\cL}{{\cal L}}
\newcommand{\gmtwo}{\ensuremath{g\!-\!2}}

\newcommand{\aj}{$D1$}
\newcommand{\ab}{$D2$}
\newcommand{\ac}{$D3$}

\newcommand{\rh}{$(\SI{1.8}{\,\giga\electronvolt})^2$}
\newcommand{\ri}{$(\SI{2.5}{\,\giga\electronvolt})^2$}
\newcommand{\rj}{$(\SI{12}{\,\giga\electronvolt})^2$}

\def\ifm#1{\relax\ifmmode#1\else$#1$\fi}  \def\plm{\ifm{\pm}}
\def\km{\kern-1.5mm}  \def\kak{\km&\km}  \def\kma{\kern-2.5mm}\def\kms{\kern-.75mm}

\begin{document}

\begin{frontmatter}

\title{An alternative evaluation of the leading-order hadronic contribution to the muon \gmtwo\ with MUonE}

%author list
\author[inst1]{Fedor Ignatov}
\ead{f.ignatov@liverpool.ac.uk}

\author[inst1]{Riccardo Nunzio Pilato}
\ead{r.pilato@liverpool.ac.uk}

\author[inst1]{Thomas Teubner}
\ead{teubner@liverpool.ac.uk}

\author[inst1,inst3]{Graziano Venanzoni}
\ead{graziano.venanzoni@liverpool.ac.uk}

\affiliation[inst1]{organization={University of Liverpool}, 
%				    addressline={Foundation Building, Brownlow Hill},
				    city={Liverpool L69 3BX},
%				    postcode={L69 3BX},
				    country={United Kingdom}}

\affiliation[inst3]{organization={INFN Sezione di Pisa}, 
	                addressline={Largo Bruno Pontecorvo 3},
	                postcode={56127},
	                city={Pisa},
	                country={Italy}}

%% Give the name of the journal
%\journal{Physics Letters B}

% The correct dates will be entered by the editor
%\date{Received: date / Accepted: date}

\begin{abstract} 
We propose an alternative method to extract the leading-order hadronic contribution to the muon \gmtwo, $a_{\mu}^\text{HLO}$, with the MUonE experiment. In contrast to the traditional method based on the integral of the hadronic contribution to the running of the electromagnetic coupling, $\Delta\alpha_{had}$, in the space-like region, our approach relies on the computation of the derivatives of $\Delta\alpha_{had}(t)$ at zero 
squared momentum transfer $t$. We show that this approach allows to extract $\sim 99\%$ of the total value of \amuhlo \ from the MUonE data, while the remaining $\sim 1\%$ can be computed combining perturbative QCD and data on $e^+e^-$ annihilation to hadrons. This 
leads to a competitive evaluation of \amuhlo \  which is robust against the parameterization used to model $\Delta\alpha_{had}(t)$ in the MUonE kinematic region, thanks to the analyticity properties of 
$\Delta\alpha_{had}(t)$, which can be expanded as a polynomial at $t\sim 0$.
\end{abstract}

%\begin{keyword}
%	muon g-2, MUonE
%\end{keyword}

\end{frontmatter}

%\linenumbers

%The article starts here:
\section{Introduction}
The muon anomalous magnetic moment, also known as the muon \gmtwo, where $g$ is the muon gyromagnetic ratio, exhibits a discrepancy between theory and experiment which persists for more than 20 years. It has received renewed interest, following the first measurement of the muon anomaly $a_\mu = (g-2)/2$  by the Muon $g\!-\!2$ Experiment at Fermilab~\cite{Muong-2:2021ojo}, subsequently confirmed by the new result with a twofold improved precision~\cite{Muong-2:2023cdq}. The comparison with the Standard Model (SM) prediction $a_\mu^\text{SM}$~\cite{Aoyama:2020ynm} is currently limited by 
tensions in the evaluation of the leading-order hadronic contribution to the muon anomaly, \amuhlo~\cite{Colangelo:2022jxc}. This term represents the main source of uncertainty of the theory prediction, due to the non-perturbative nature of QCD at low energy. A recent computation of \amuhlo \ based on lattice QCD, performed by the BMW Collaboration \cite{Borsanyi:2020mff}, indeed shows a $2.1\sigma$ tension with the one used in the SM evaluation of $a_\mu$ \cite{Aoyama:2020ynm}, which is based on a data-driven approach involving data for $e^+e^- \rightarrow \text{hadrons}$ cross sections. Moreover, a new experimental measurement of $e^+e^- \rightarrow \pi^+\pi^-$ channel from the CMD-3 experiment disagrees with the previous measurements \cite{CMD-3:2023alj}.
New calculations from other lattice QCD groups and new results from $e^+e^-$ colliders are expected to shed light on these tensions in the next few years \cite{Colangelo:2022jxc}. 
\newline
Recently a new approach has been proposed to compute \amuhlo, based on the measurement of the hadronic contribution to the running of the electromagnetic coupling, $\Delta\alpha_{had}$, in the space-like region \cite{MUonEApproach}.
The elastic scattering of high-energy muons on atomic electrons has been identified as an ideal process for this measurement and an experimental proposal, called MUonE, has been put forward at CERN to extract $\Delta\alpha_{had}$ from a precise measurement of the shape of the $\mu^+ e^- \rightarrow \mu^+ e^-$ elastic process  \cite{MUonEExperiment}. The goal of MUonE is to determine \amuhlo \ with a $\sim 0.3\%$ statistical and a comparable systematic uncertainty, using the following integral \cite{LautrupDeRafael}:
\begin{equation}
	a_\mu^\text{HLO} = \frac{\alpha}{\pi}\int_0^1 dx(1-x)\Delta\alpha_{had}[t(x)], \qquad t(x) = \frac{x^2m_\mu^2}{x-1} < 0,
	\label{eq:eqMUonEIntegral}
\end{equation}
where $\alpha$ is the fine structure constant, $m_\mu$ is the muon mass, and $t$ is the space-like squared momentum transfer. The \mbox{160 GeV} muon beam available at the M2 beamline at CERN allows to cover directly the momentum transfer range \mbox{$-\SI{0.153}{\,\giga\electronvolt^2} < t < -\SI{0.001}{\,\giga\electronvolt^2}$}, which is equivalent to $0.258 < x < 0.936$. This corresponds to $\sim 86\%$ of the integral in Eq.~\ref{eq:eqMUonEIntegral}, while the remaining fraction can be obtained by extrapolating \dalphahad \ outside the MUonE region by an appropriate analytical function or alternatively by using lattice data. 
In the first case the space-like integral of Eq.~\ref{eq:eqMUonEIntegral}  is sensitive to the behaviour of the parameterization chosen to model \dalphahad \ in the whole $t$-region, particularly in the asymptotic limit $t \rightarrow -\infty$, which could affect the extraction of \amuhlo.\footnote{A convenient choice based on the analytic formula for the leading-order QED contribution to the running of $\alpha$ in the space-like region allows to compute \amuhlo \ at the required level of precision with negligible bias when time-like data are used as input~\cite{Abbiendi_2022} (see also Section \ref{sec3}).}
 In the following, we will discuss a different approach based on the evaluation of the derivatives of \dalphahad \ at zero momentum transfer. This leads to an evaluation of \amuhlo \ which is rather insensitive to the functional form adopted to describe the behaviour of \dalphahad \ and will provide an alternative and competitive way to determine \amuhlo \ with MUonE.

\section{Description of the method}
\begin{figure}[h!]
    \centering
    \includegraphics[width=7.5cm,keepaspectratio]{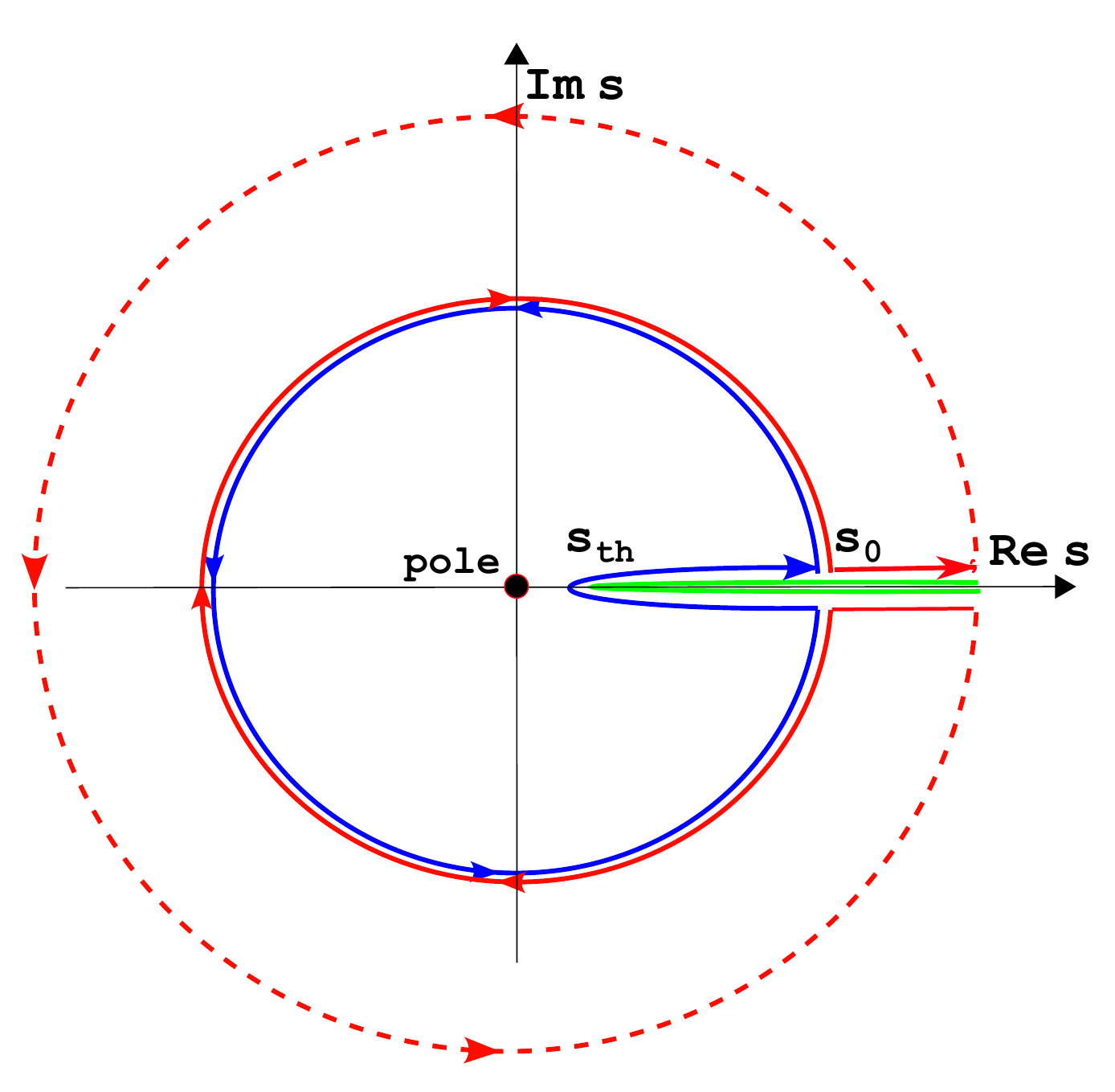}
    \caption{Blue (red): closed path in the complex $s$-plane used to calculate the contour integral in Eq.~\ref{eq:eqLowEnergyCauchy} (Eq.~\ref{eq:eqHighEnergyCauchy}).}
    \label{fig:figCauchyPath}
\end{figure}

\begin{figure}[h]
    \centering
    \includegraphics[width=14.5cm,keepaspectratio]{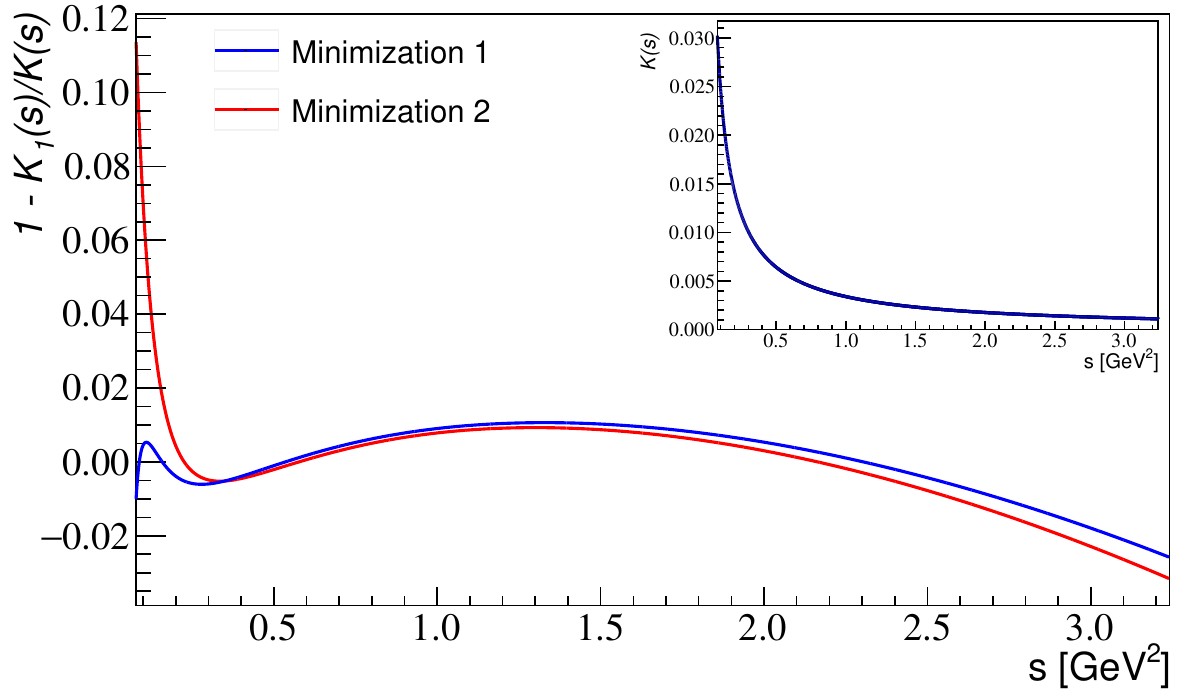}
    \caption{Fractional difference between the approximated kernel $K_1(s)$ and the exact kernel $K(s)$ for the two minimization methods for $s_0 = (\SI{1.8}{\giga\electronvolt})^2$. The insertion shows the exact kernel in the same range.}
    \label{fig:figK1}
\end{figure}

The new method is mainly based on \cite{Dominguez12, Dominguez17}, where different quark flavours have been treated separately, and the dominant light-quark contributions to the hadronic vacuum polarization function $\Pi_{had}(s)$ have been computed either through a model-dependent approach \cite{Dominguez12}, or using lattice QCD calculations \cite{Dominguez17}. The same strategy is not applicable to MUonE, since MUonE will provide an inclusive measurement of $\Pi_{had}(t)$ containing contributions from all the quark flavours.
In the following, we summarize the original procedure and describe how it can be adapted to the MUonE case.
\noindent We start from the well known dispersion relation \cite{timelike1, timelike2, timelike2_erratum, timelike3}
\begin{equation}
    a_\mu^\text{HLO} = \frac{\alpha^2}{3\pi^2}\int_{s_\text{th}}^{\infty}\frac{ds}{s}K(s)R(s),
    \label{eq:eqTimeLike}
\end{equation}
where, due to contributions from the $\pi^0\gamma$ channel, the threshold $s_\text{th}$ is usually identified as $m_{\pi^0}^2$, with $m_{\pi^0}$ being the $\pi^0$ mass. The kernel function $K(s)$ is given by
\begin{equation}
	K(s) = \int_{0}^{1}dx \frac{x^2(1-x)}{x^2 + \frac{s}{m_\mu^2}(1-x)},
    \label{eq:eqKernelFunction}
\end{equation}
and $R(s)$ is the ratio of the bare (i.e.\ excluding vacuum polarization effects) \mbox{$e^+e^- \rightarrow \text{hadrons}$} annihilation cross section to the Born \mbox{$e^+e^- \rightarrow \mu^+\mu^-$} pointlike one, $\sigma_{pt}= 4\pi\alpha^2/(3s)$. Taking advantage of the optical theorem, $R(s)$  can be related to the imaginary part of the hadronic vacuum polarization function $\Pi_{had}(s)$ as follows:
\begin{equation}
	-\text{Im}\Pi_{had}(s) = \frac{\alpha}{3}R(s).
    \label{eq:eqOpticalTheorem}
\end{equation}
\newline The dispersive integral in Eq.~\ref{eq:eqTimeLike} can be split in two terms at a given value $s_0$, above which $R(s)$ can be safely computed using perturbative QCD (pQCD). As originally proposed in \cite{Dominguez12}\footnote{Different approaches to evaluate the dispersive integral by an approximate kernel function have been discussed in~\cite{Davier:1998si,Groote:1998pk}.}, it is convenient to approximate $K(s)$ by a meromorphic function $K_1(s)$ for $s \leq s_0$,
\begin{equation}
    K_1(s) = c_0s + \sum_{n = 1}^3 \frac{c_n}{s^n},
    \label{eq:eqK1}
\end{equation}
and the low energy part of the dispersive integral can be written as
\begin{equation}
    -\frac{\alpha}{\pi} \int_{s_\text{th}}^{s_0} \frac{ds}{s} K(s)\frac{\text{Im}\Pi_{had}(s)}{\pi} \ =\ -\frac{\alpha}{\pi} \left[\int_{s_\text{th}}^{s_0} \frac{ds}{s} [K(s) - K_1(s)]\frac{\text{Im}\Pi_{had}(s)}{\pi} \ +\  \int_{s_\text{th}}^{s_0} \frac{ds}{s} K_1(s)\frac{\text{Im}\Pi_{had}(s)}{\pi}\right],
\label{eq:eqLowEnergyIntegral}
\end{equation}
where Eq.~\ref{eq:eqOpticalTheorem} was used to express $R(s)$ in terms of the hadronic vacuum polarization function. Cauchy's theorem can then be employed to handle the second term on the right-hand side \cite{Dominguez12, Dominguez17}:
\begin{equation}
	\int_{s_\text{th}}^{s_0}\frac{ds}{s}K_1(s)\frac{\text{Im}\Pi_{had}(s)}{\pi} \ =\ \text{Res}\left[\Pi_{had}(s)\frac{K_1(s)}{s}\right]_{s=0} \ -\  \frac{1}{2\pi i}\oint_{|s| = s_0}\frac{ds}{s}K_1(s)\Pi_{had}(s)\bigg|_{\text{pQCD}}.
 \label{eq:eqLowEnergyCauchy}
\end{equation}
Here, the contour integral around the circle of radius $s_0$ can be calculated using pQCD to evaluate the hadronic vacuum polarization function, whereas the residual can be written in terms of derivatives of $\Pi_{had}(s)$ at zero momentum transfer. Exploiting the functional form of the approximated kernel in Eq.~\ref{eq:eqK1},
\begin{equation}
    \text{Res}\left[\Pi_{had}(s)\frac{K_1(s)}{s}\right]_{s=0} \ =\  \sum_{n = 1}^3 \frac{c_n}{n!} \frac{d^{(n)}}{ds^n} \Pi_{had}(s)\bigg|_{s = 0} \ =\  \sum_{n = 1}^3 \frac{c_n}{n!} \frac{d^{(n)}}{dt^n} \Delta\alpha_{had}(t)\bigg|_{t = 0},
    \label{eq:eqResidualsTerm}
\end{equation}
where the analyticity of $\Pi_{had}$ and its derivatives at zero momentum transfer has been exploited to move the evaluation of the hadronic vacuum polarization from positive to negative squared momentum transfer. The relation $\Delta\alpha_{had}(t) = \text{Re}\Pi_{had}(t)$ has been used in the last step to link the hadronic vacuum polarization function to the hadronic contribution to the running of $\alpha$.
\newline The high energy tail of the dispersive integral can be treated in a similar way:
\begin{equation}
    -\frac{\alpha}{\pi}\int_{s_0}^\infty \frac{ds}{s}K(s) \frac{\text{Im}\Pi_{had}(s)}{\pi} \ =\  -\frac{\alpha}{\pi} \left[\int_{s_0}^\infty \frac{ds}{s} [K(s) - \Tilde{K}_1(s)]\frac{\text{Im}\Pi_{had}(s)}{\pi} \ +\  \int_{s_0}^\infty \frac{ds}{s} \Tilde{K}_1(s)\frac{\text{Im}\Pi_{had}(s)}{\pi}\right], \label{eq:eqHighEnergyIntegral}
\end{equation}
where $\Tilde{K}_1(s) = K_1(s) - c_0s$. Following the same technique implemented for the low energy component, Cauchy's theorem can be applied to the second integral on the right hand side of Eq.~\ref{eq:eqHighEnergyIntegral}, using the red closed path shown in Fig.~\ref{fig:figCauchyPath}. In this case, the integrand is free of poles and the contour integral over $s$ with radius $|s| \rightarrow \infty$ is vanishing. This leads to
\looseness=-1
\begin{equation}
    \int_{s_0}^\infty \frac{ds}{s} \Tilde{K}_1(s)\frac{\text{Im}\Pi_{had}(s)}{\pi} \ =\  \frac{1}{2\pi i}\oint_{|s| = s_0}\frac{ds}{s}\Tilde{K}_1(s)\Pi_{had}(s)\bigg|_{\text{pQCD}}.
    \label{eq:eqHighEnergyCauchy}
\end{equation}
\newline Rearranging Eqs.~\ref{eq:eqLowEnergyIntegral}, \ref{eq:eqLowEnergyCauchy},  \ref{eq:eqHighEnergyIntegral} and \ref{eq:eqHighEnergyCauchy}, \amuhlo \ can be calculated as the sum of four terms 
\begin{equation}
    a_\mu^\text{HLO} \ =\  a_\mu^{\text{HLO (I)}} + a_\mu^{\text{HLO (II)}} + a_\mu^{\text{HLO (III)}} + a_\mu^{\text{HLO (IV)}},
    \label{eq:eqAmuHLODer}
\end{equation}
where
\begin{align}
a_\mu^{\text{HLO (I)}} &\ =\  -\frac{\alpha}{\pi} \sum_{n = 1}^3 \frac{c_n}{n!} \frac{d^{(n)}}{dt^n} \Delta\alpha_{had}(t)\bigg|_{t = 0}, \label{eq1}\\
a_\mu^{\text{HLO (II)}} &\ =\ \frac{\alpha}{\pi} \frac{1}{2\pi i}\oint_{|s| = s_0}\frac{ds}{s}c_0s \ \Pi_{had}(s)\bigg|_{\text{pQCD}},\\
a_\mu^{\text{HLO (III)}} &\ =\ \frac{\alpha^2}{3\pi^2} \int_{s_\text{th}}^{s_0} \frac{ds}{s} [K(s) - K_1(s)]R(s),\\
a_\mu^{\text{HLO (IV)}} &\ =\ \frac{\alpha^2}{3\pi^2} \int_{s_0}^\infty \frac{ds}{s} [K(s) - \Tilde{K}_1(s)]R(s).
\end{align}
$a_\mu^{\text{HLO (I)}}$ will be computed using MUonE data. 
This term represents $\sim99\%$ of the total value of \amuhlo, as will be shown in the following. The other three terms contribute to the remaining $\sim1\%$.
$a_\mu^{\text{HLO (II)}}$ will be calculated via pQCD, $a_\mu^{\text{HLO (III)}}$ with $e^+e^-$ data, while both $e^+e^-$ data and pQCD will be used to compute $a_\mu^{\text{HLO (IV)}}$. 
\newline In the following sections, the calculation of the different contributions will be discussed in detail. The robustness of the proposed method will be tested using three different values of $s_0$: $(\SI{1.8}{\,\giga\electronvolt})^2$, $(\SI{2.5}{\,\giga\electronvolt})^2$ and $(\SI{12}{\,\giga\electronvolt})^2$. Two different techniques will be used to determine the coefficients of $K_1(s)$. The first (called {\it Minimization 1}) consists of a least squares minimization of the difference between the approximated and the analytical kernel. The second ({\it Minimization 2}) aims at minimizing the contribution of $e^+e^-$ data in the calculation of \amuhlo, and is carried out by fixing the coefficient $c_3$ to be 1/2 of its value obtained from {\it Minimization 1}, then minimizing $\int_{s_\text{th}}^{s_0} \frac{ds}{s} |K(s) - K_1(s)|R(s)$ to find the other coefficients. Figure \ref{fig:figK1} shows the fractional difference between $K_1(s)$ and the exact kernel for both the minimization techniques and the choice $s_0 = (\SI{1.8}{\giga\electronvolt})^2$. Table~\ref{tab:0} shows the coefficients for the two minimizations. Both methods provide a good approximation of $K(s)$, with a different sensitivity to the third derivative $\frac{d^3\Delta\alpha_{had}(t)}{dt^3}$  at $ t=0$ (which will be shown to have the largest uncertainty, see Table~\ref{tab:1}).
\begin{table}[h]
    \centering
    \begin{tabular}{ccccccc}
    \hline\hline
    & \multicolumn{3}{c}{Minimization 1} &  \multicolumn{3}{c}{Minimization 2}   
    \\
    \hline
       Coefficients  & $(\SI{1.8}{\,\giga\electronvolt})^2$ & $(\SI{2.5}{\,\giga\electronvolt})^2$ &  $(\SI{12}{\,\giga\electronvolt})^2$ & $(\SI{1.8}{\,\giga\electronvolt})^2$ & $(\SI{2.5}{\,\giga\electronvolt})^2$ &  $(\SI{12}{\,\giga\electronvolt})^2$\\
       \hline
       $c_0\cdot10^5$ & 2.206 & 0.7326 & 0.002164 & 2.419 & 1.011 & 0.003743 \\
       $c_1\cdot10^3$ & 3.486 & 3.512 & 3.555 & 3.482 & 3.494 & 3.520 \\
       $c_2\cdot10^4$ & -1.484 & -1.554 &  -1.684 & -1.402 & -1.443 & -1.564 \\
       $c_3\cdot10^6$ & 4.869 & 5.294 & 6.128 & 2.434 & 2.647 & 3.064 \\
       \hline\hline
    \end{tabular}
    \caption{Coefficients of the approximated kernel function $K_1(s)$ for the two minimizations and the three choices of $s_0$.}
    \label{tab:0}
\end{table}

\section{Evaluation of $a_\mu^{\text{HLO (I)}}$}
\label{sec3}
\begin{figure}[h]
    \centering
    \includegraphics[width=0.49\textwidth, keepaspectratio]{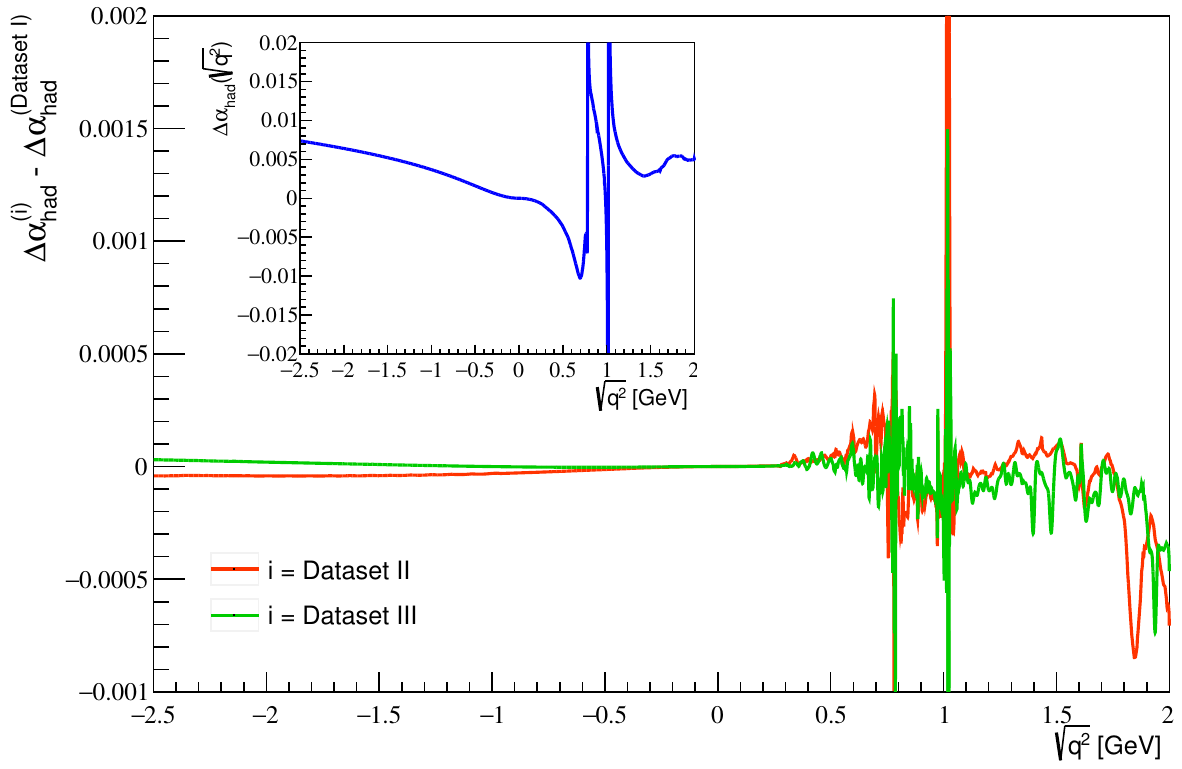}
    \includegraphics[width=0.49\textwidth, keepaspectratio]{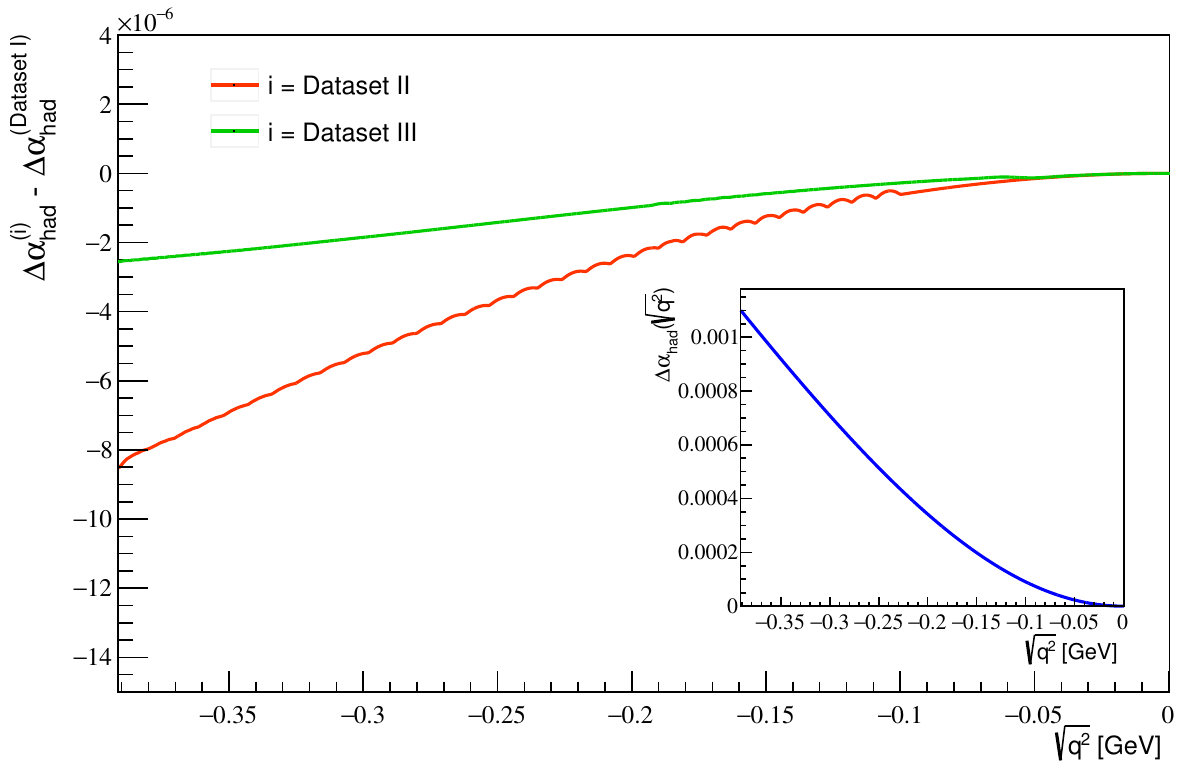}
    \caption{Difference of data compilation for $\Delta\alpha_{had}$ of Dataset II and III respect to Dataset I in the range  $-2.5< \sqrt{q^2}< 2$ GeV (Left) and for the MUonE kinematic range (Right). The insertions show the behaviour of $\Delta\alpha_{had}$ computed using  Dataset I.}
    \label{fig:figDAHad_dataInput}
\end{figure}

The first term $a_\mu^{\text{HLO (I)}}$ in Eq.~\ref{eq:eqAmuHLODer} depends on the derivatives of \dalphahad \ at zero momentum transfer, which can be obtained by fitting the MUonE data with a convenient functional form.
Since space-like data in the MUonE range are not available, we use different data compilations of \dalphahad \  obtained from $e^+e^-$ data in the time-like region using the dispersive integral:
\begin{equation}
    \Delta\alpha_{had}(q^2) = - \frac{\alpha}{3\pi} q^2 \,{\rm P}\int_{s_{th}}^\infty ds \frac{R(s)}{s(s-q^2)},
    \label{disint}
\end{equation}
where $q^2$ is a generic squared momentum transfer and P denotes the principal value of the integral. The data compilations produced in \cite{Fedor}, \cite{FJ} and \cite{KNT}, indicated respectively as Dataset I, II and III, will be used.
%Figure~\ref{fig:figDAHad_dataInput} represents $\Delta\alpha_{had}(q^2)$ computed from the three compilations both for positive and negative squared momentum transfer. The small deviations between the three compilations are due to different strategies adopted in combining data from various experiments. The influence of the hadronic resonances on the vacuum polarization effects is evident for positive squared momentum transfer, whereas $\Delta\alpha_{had}(q^2)$  is a smooth function for $q^2 < 0$.
%The three compilations adopted different strategies to combine data from various experiments, leading to 
Figure \ref{fig:figDAHad_dataInput} shows the difference between the three compilations. Deviations from 0 are due to different strategies adopted in combining data from various experiments. The influence of the hadronic resonaces on the vacuum polarization effects is evident for positive squared momentum transfer (Figure \ref{fig:figDAHad_dataInput}, Left), whereas $\Delta\alpha_{had}(q^2)$  is a smooth function for $q^2 < 0$, as shown in Figure \ref{fig:figDAHad_dataInput}, Right.
\par To assess the sensitivity of our method to the parameterization adopted for \dalphahad, we tested several possible parameterizations which are capable of reproducing the behaviour of \dalphahad \ in the MUonE kinematic range:
\begin{enumerate}
    \item ``Lepton Like'' (LL) parameterization. This is inspired by the one-loop QED result for the vacuum polarisation induced by a lepton pair in the space-like region, and is also used in MUonE to calculate \amuhlo \ through the space-like integral in Eq.~\ref{eq:eqMUonEIntegral} \cite{Abbiendi_2022}:
    \begin{equation}
        \Delta\alpha_\text{had}(t) = K M \left( -\frac{5}{9}-\frac{4 M}{3 t} +\left(\frac{4 M^2}{3 t^2}+\frac{M}{3 t} -\frac{1}{6}\right) \frac{2}{\sqrt{1-\frac{4 M}{t}}} \log \left|\frac{1-\sqrt{1-\frac{4
        M}{t}}}{1+\sqrt{1-\frac{4 M}{t}}}\right| \right),
        \label{DalphaLLparam}
    \end{equation}
 where $M$ is the squared lepton mass and $K = \alpha/(\pi M)$ in the leptonic case. On the contrary, these two parameters do not have a precise physics interpretation when modeling the hadronic running since \dalphahad \ is not calculable in perturbation theory.
In the limit of very small $t$, the LL function is approximated by the following expansion:
    \begin{equation}
        \Delta\alpha_\text{had}(t) = - K \left( \frac{t}{15} + \frac{t^2}{140 M} + \frac{t^3}{945 M^2}+O(t^4)\right),
        %\frac{K t}{15}-\frac{K t^2}{140 M}-\frac{K t^3}{945 M^2}-\frac{K t^4}{5544 M^3}-\frac{K t^5}{30030 M^4}+O(t)^7,
        %\frac{1}{15} \frac{k}{M} t +....,
        \label{DalphaLLlowt}
    \end{equation}
    which corresponds to the dominant behaviour in the MUonE kinematic range. The first three derivatives at zero momentum transfer are:
    \begin{equation*}
        D1 = \frac{d\Delta\alpha_\text{had}(t)}{dt}\bigg|_{t = 0} = -\frac{K}{15}, \qquad
        D2 = \frac{d^2\Delta\alpha_\text{had}(t)}{dt^2}\bigg|_{t = 0} = - \frac{K}{70 M}, \qquad 
        D3 = \frac{d^3\Delta\alpha_\text{had}(t)}{dt^3}\bigg|_{t = 0} =  - \frac{2 K}{315 M^2}.
    \end{equation*}

    \item A Pad\'e approximant (Pad\'e parameterization) with three free parameters $P_{1,2,3}$:
    \begin{equation}
    \Delta\alpha_\text{had}(t) = P_1 t \frac{1 + P_2 t}{1 + P_3 t}. 
    \label{pade}
    \end{equation}
    The expansion in the limit of very small $t$ gives the following expression:
    \begin{equation}
    \begin{split}
    \Delta\alpha_\text{had}(t) = P_1 t + P_1(P_2 -P_3) t^2 + P_1 P_3(-P_2 +P_3) t^3 + O(t^4),
    \end{split}
    \label{DalphaLLlowt_pade}
    \end{equation}
    with the first three derivatives at zero momentum transfer:
    \begin{equation*}
        D1 = P_1, \qquad D2 = 2P_1(P_2 - P_3), \qquad D3 = - 6P_1 P_3(P_2 - P_3).
    \end{equation*}
    
    \item A third order polynomial (Pol parameterization) with three free parameters $P_{1,2,3}$. Since $\Delta\alpha_{had}(t = 0) = 0$, the $P_0$ term is fixed to be zero:
    \begin{equation}
    \Delta\alpha_\text{had}(t) = P_1 t+P_2 t^2+P_3t^3,
    \label{pol}
    \end{equation}
    which gives the following derivatives:
    \begin{equation*}
        D1 = P_1, \qquad  D2 = 2P_2, \qquad D3 = 6P_3. 
    \end{equation*}
    
    \item ``Reconstruction Approximants'' (GdR parameterization) \cite{GdR2022}.
    This is based on the analytic properties of the hadronic vacuum polarization function $\Pi_{had}(t)$ in QCD and takes the form ~\cite{GdR2022}:
    \begin{align}
        &\Delta\alpha_\text{had}(t) = \sum_{n=1}^{\rm N} \cA(n,{\rm L})   \left(\frac{\sqrt{1-\frac{t}{t_0}}-1}{\sqrt{1-\frac{t}{t_0}}+1}  \right)^n + \sum_{p=1}^{\left\lfloor \frac{{\rm L}+1}{2}\right\rfloor}\cB(2p-1)\ {\rm Li}_{2p-1}\left(\frac{\sqrt{1-\frac{t}{t_0}}-1}{\sqrt{1-\frac{t}{t_0}}+1}  \right), 
    \end{align}
    where $\cA(n,{\rm L})$ and $\cB(2p-1)$ are free parameters which can be constrained by theory \cite{GdR2022}, $t_0$ is an energy scale (set in this work to $4 m^2_{\pi^{\pm}}$), and ${\rm Li}_{2p-1}$ are Polylogarithms of degree $2p-1$. In the following, we limit the expansion to the case L=1, N=3: 
    \begin{equation}
    \Delta\alpha_\text{had}(t) = A_1 \cS_1 \ + \ A_2 \cS_2 \ + \ A_3 \cS_3 \ + \ B_1 \cL_1,
    \end{equation}
    where
    \begin{eqnarray*}
     \cS_i &= &\left(\frac{\sqrt{1 -\frac{t}{t_0}}-1}{\sqrt{1-\frac{t}{t_0}}+1}\right)^i, \qquad \quad A_i =  \cA(i,1) \quad {\rm for}\ \ i=1,\ 2,\ 3, \\
       \cL_1 & = & {\rm Li}_{1}\left(\frac{\sqrt{1 -\frac{t}{t_0}}-1}{\sqrt{1-\frac{t}{t_0}}+1}\right), \qquad \quad    B_1  =  \cB(1). 
    \end{eqnarray*}
    In the limit of small momentum transfer, the GdR function can be approximated by:
    \begin{eqnarray*}
    \Delta\alpha_\text{had}(t) =  -\frac{(A_1+B_1) t}{4 t_0}-\frac{(4A_1-2A_2+3B_1) t^2}{32 t_0^2}-\frac{\left(3 (5 A_1-4 A_2+A_3)+10 B_1\right) t^3}{192 t_0^3}+O(t^4).
    \end{eqnarray*}
    The first derivatives at zero momentum transfer can be computed as:
    \begin{equation*}
    D1 = -\frac{A_1 + B_1}{4t_0}, \qquad D2 = \frac{-4A_1 + 2A_2 - 3B_1}{16t_0^2}, \qquad D3 = \frac{-15A_1 + 12A_2 - 3A_3 - 10B_1}{32t_0^3}. 
    \end{equation*}
    Five different variants of this parameterization have been considered:
    \begin{enumerate}
        \item GdR1: $\Delta\alpha_\text{had}(t) = A_1 \cS_1 + B_1 \cL_1$,
        where $A_1$ is a free parameter and $B_1$ is constrained to its six-quark asymptotic freedom value: $B_1 = 2\frac{\alpha}{\pi}\frac{5}{3}=0.00774273$ \cite{GdR2022}.
        \item GdR2: $\Delta\alpha_\text{had}(t) = A_1 \cS_1 + B_1 \cL_1$,
        where $A_1$ and $B_1$ are free parameters.
        \item GdR3: $\Delta\alpha_\text{had}(t) = A_1 \cS_1+A_2 \cS_2 + B_1 \cL_1$, where $A_1, A_2$ are free parameters and $B_1$ is constrained to its six-quark asymptotic freedom value.
        \item GdR4: $\Delta\alpha_\text{had}(t) = A_1 \cS_1+A_2 \cS_2 + A_3 \cS_3 + B_1 \cL_1$, where $A_1, A_2$ are free parameters, $A_3$ is constrained to $A_3=(2A_2-A_1-B/2)/3$ \cite{GdR2022} and $B_1$ is constrained to its six-quark asymptotic freedom value.
        \item GdR5: $\Delta\alpha_\text{had}(t) = A_1 \cS_1+A_2 \cS_2 + A_3 \cS_3 + B_1 \cL_1$, where $A_1, A_2, A_3$ are free parameters and $B_1$ is constrained to its six-quark asymptotic freedom value.
    \end{enumerate}
    
\end{enumerate}

\par
\begin{figure}[h]
    \centering
    \includegraphics[width=0.48\textwidth, keepaspectratio]{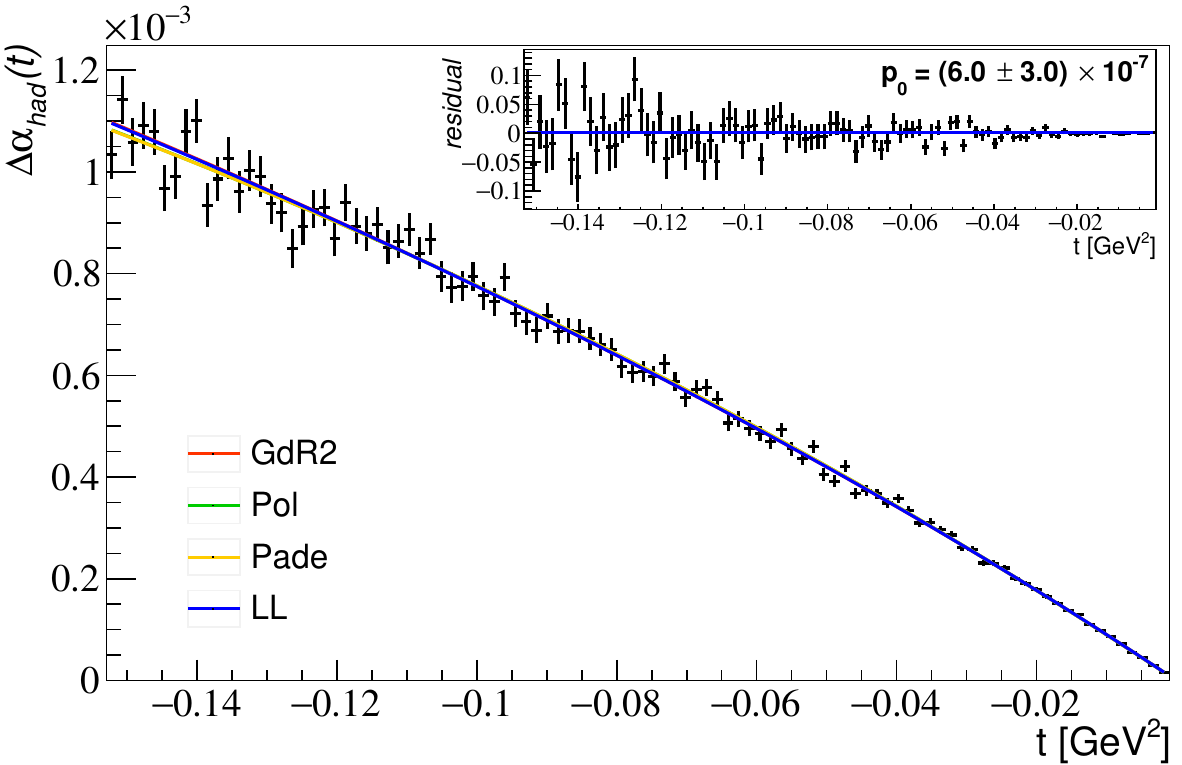}
    \includegraphics[width=0.49\textwidth, keepaspectratio]{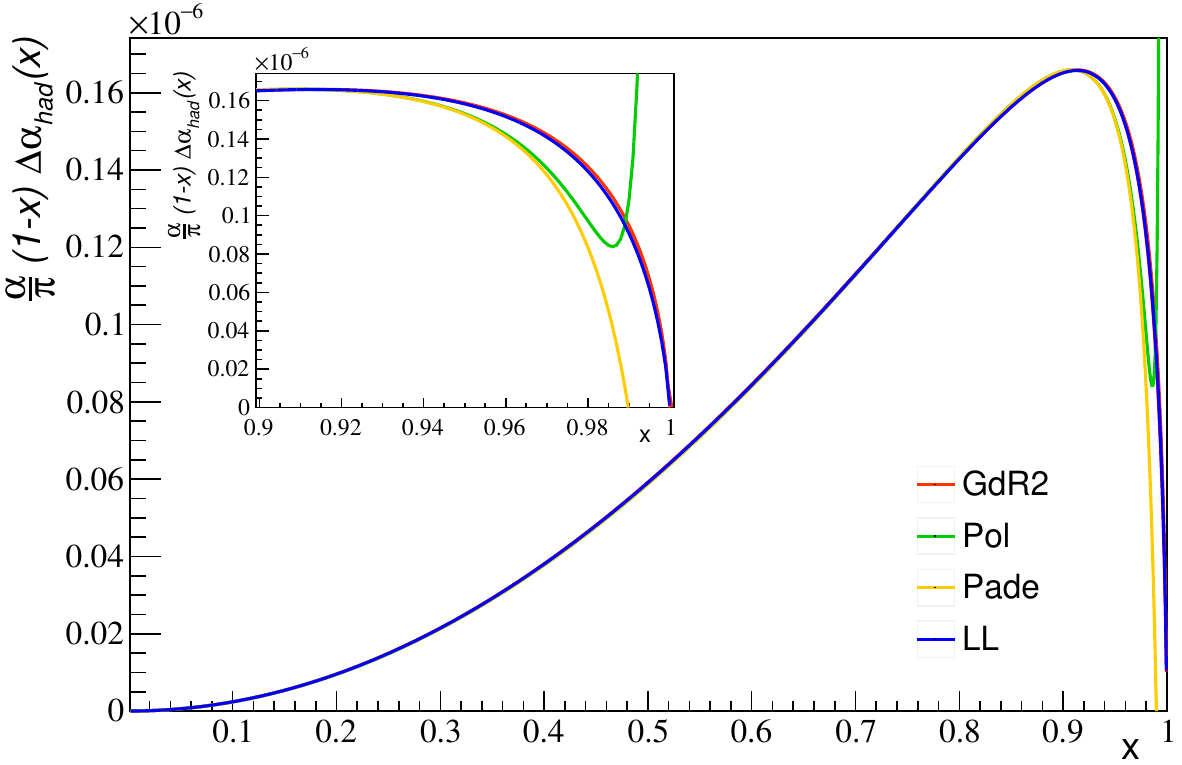}
    \caption{Left: example of fit results for a few parameterizations of \dalphahad. The insertion shows the fit residuals for the LL best fit of the same pseudo-experiment. Right: integrand of the MUonE integral in Eq.~\ref{eq:eqMUonEIntegral}, computed using different parameterizations fitted to the same pseudo-experiment. The insertion zooms in on the kinematic range not accessible to MUonE. Note that the Pad\'e approximant fails to follow the expected behaviour, while the third order polynomial diverges for $x\rightarrow 1$ ($t\rightarrow-\infty$).}
    \label{fig:figFitResults}
\end{figure}

 For each \dalphahad \ compilation, $10^4$ pseudo-experiments were simulated in the MUonE momentum transfer region -0.153 GeV$^2<t<-0.001$ GeV$^2$. The higher limit is needed to reproduce the geometric acceptance of MUonE, which will include all the elastic events where the electron angle is $< \SI{32}{\milli\radian}$. Statistical fluctuations have been added independently to each \dalphahad \ point according to the planned full integrated luminosity of MUonE, $\SI{15}{\femto\barn^{-1}}$ \cite{MUonEExperiment}. A $\chi^2$ fit has been performed for each pseudo-experiment using all the parameterizations described above, and for each iteration both the values of \amuhlo \ (according to Eq.~\ref{eq:eqMUonEIntegral}) and of $a_\mu^{\text{HLO (I)}}$ were calculated. In particular, $a_\mu^{\text{HLO (I)}}$ was calculated for the three choices of $s_0$ and for the two minimizations used to determine the approximated kernel $K_1(s)$. This allowed to test the robustness of the proposed method extensively.
\newline As an example, Fig.~\ref{fig:figFitResults}, Left, shows the fit results of a few parameterizations of \dalphahad \ for a given pseudo-experiment generated with Dataset I. All the parameterizations describe the MUonE simulated data well. However, differences arise outside the MUonE experimental range, as is demonstrated in Fig.~\ref{fig:figFitResults}, Right, which shows the integrand in Eq.~\ref{eq:eqMUonEIntegral}. In particular, the Pad\'e approximant and the third order polynomial fail to describe \dalphahad \ for $x\rightarrow 1$, which corresponds to $t \rightarrow -\infty$. As a consequence, values of \amuhlo \ computed from the integral in Eq.~\ref{eq:eqMUonEIntegral} using these two approximations will not be in agreement with the expected (input) value. On the other hand, the results of the three derivatives are quite stable, as is shown in Table~\ref{tab:1}, which compares the results of the three derivatives for all the different parameterizations used and the three \dalphahad \ compilations. These values must be compared with the ones obtained by the corresponding dispersive integrals calculated directly from the time-like data, using Dataset I:  $D1 = (-9.24 \pm 0.04)\cdot 10^{-3}$, $D2 = (-3.86 \pm 0.02)\cdot 10^{-2}$, $D3 = (-3.90 \pm 0.03)\cdot 10^{-1}$.
Table~\ref{tab:2} shows the values of $a_\mu^{\text{HLO (I)}}$ for the different parameterizations and different $s_0$ values for the two minimizations using Eq.~\ref{eq1} with Dataset I. As can be seen, the results are quite stable, independent of the parameterization used. This is a consequence of the analyticity of the hadronic vacuum polarization function $\Pi_{had}(t)$ which can be approximated by a polynomial through a Taylor expansion of \dalphahad\ at $t\sim0$ using Eq.~\ref{disint}.
%%%TABLE 1
\begin{table}[h!]
\vspace{1cm}
\renewcommand{\arraystretch}{.9}
\begin{center}
\begin{tabular}{ccccc}
\hline\hline
 \multicolumn{2}{c}{ }  & \multicolumn{3}{c}{Dataset}  \\
\hline
Parameterization & Values & I & II & III \\
\hline
\kak\aj\kak-9.23\plm0.05\kak-9.17\plm0.05\kak-9.20\plm0.05\km\\ 
\multirow{1}{0pt}{LL}\kak\ab\kak-3.76\plm0.22\kak-3.77\plm0.22\kak-3.73\plm0.22\km\\ 
\kak\ac\kak-3.19\plm0.35\kak-3.21\plm0.35\kak-3.14\plm0.35\km\\ 
\hline
\kak\aj\kak -9.24\plm0.08\kak-9.18\plm0.08\kak-9.20\plm0.08\km\\ 
\multirow{1}{0pt}{Pad\'e}\kak\ab\kak-3.81\plm0.62\kak-3.81\plm0.62\kak-3.75\plm0.61\km\\ 
\kak\ac\kak-2.14\plm2.42\kak-2.21\plm2.41\kak-2.03\plm2.42\km\\ 
\hline
\kak\aj\kak-9.22\plm0.07\kak-9.16\plm0.07\kak-9.19\plm0.07\km\\ 
\multirow{1}{0pt}{Pol}\kak\ab\kak-3.55\plm0.45\kak-3.56\plm0.45\kak-3.50\plm0.44\km\\ 
\kak\ac\kak-1.83\plm0.89\kak-1.84\plm0.88\kak-1.79\plm0.88\km\\ 
\hline
\kak\aj\kak-9.26\plm0.03\kak-9.19\plm0.03\kak-9.23\plm0.03\km\\ 
\multirow{1}{0pt}{GdR1}\kak\ab\kak-3.92\plm0.04\kak-3.82\plm0.04\kak-3.88\plm0.04\km\\ 
\kak\ac\kak-3.03\plm0.10\kak-2.80\plm0.10\kak-2.93\plm0.10\km\\ 
\hline
\kak\aj\kak-9.23\plm0.06\kak-9.18\plm0.06\kak-9.20\plm0.06\km\\ 
\multirow{1}{0pt}{GdR2}\kak\ab\kak-3.73\plm0.31\kak-3.75\plm0.31\kak-3.69\plm0.31\km\\ 
\kak\ac\kak-2.46\plm0.94\kak-2.58\plm0.94\kak-2.33\plm0.95\km\\ 
\hline
\kak\aj\kak-9.23\plm0.06\kak-9.18\plm0.06\kak-9.20\plm0.06\km\\ 
\multirow{1}{0pt}{GdR3}\kak\ab\kak-3.70\plm0.36\kak-3.74\plm0.36\kak-3.65\plm0.36\km\\ 
\kak\ac\kak-2.25\plm1.30\kak-2.47\plm1.31\kak-2.13\plm1.29\km\\ 
\hline
\kak\aj\kak-9.24\plm0.06\kak-9.19\plm0.06\kak-9.21\plm0.06\km\\ 
\multirow{1}{0pt}{GdR4}\kak\ab\kak-3.89\plm0.31\kak-3.91\plm0.32\kak-3.84\plm0.32\km\\ 
\kak\ac\kak-2.95\plm1.11\kak-3.14\plm1.11\kak-2.84\plm1.11\km\\ 
\hline
\kak\aj\kak-9.23\plm0.09\kak-9.18\plm0.09\kak-9.20\plm0.09\km\\ 
\multirow{1}{0pt}{GdR5}\kak\ab\kak-3.74\plm1.23\kak-3.75\plm1.23\kak-3.64\plm1.24\km\\ 
\kak\ac\kak-2.90\plm3.23\kak-2.91\plm3.37\kak-1.80\plm4.52\km\\ 
\hline\hline
\end{tabular}\vglue2mm
\end{center}
\caption{Values of the first three derivatives of \dalphahad \ at zero momentum transfer for the parameterizations considered in this study and the three different datasets. $D1$ is given in units of $10^{-3}$, $D2$ in units of $10^{-2}$, and $D3$ in units of $10^{-1}$.}
\label{tab:1}
\end{table}

%%%TABLE 2
\begin{table}[h!]
\renewcommand{\arraystretch}{.9}
\begin{tabular}{ccccccccc}
\hline\hline
Minimization I & \multicolumn{8}{c}{\amud}  \\
\hline
$s_0$ values & LL & Pad\'e & Pol & GdR1 & GdR2 & GdR3 & GdR4 & GdR5 \\
\hline\rh\kak688.7\plm2.2\kak688.7\plm2.9\kak688.9\plm2.9\kak688.2\plm2.2\kak688.0\plm2.2\kak688.0\plm2.2\kak687.0\plm2.3\kak688.0\plm2.6\km\\ 
\hline\ri\kak691.7\plm2.2\kak691.6\plm3.0\kak691.8\plm3.0\kak691.0\plm2.2\kak690.8\plm2.2\kak690.8\plm2.2\kak689.8\plm2.3\kak690.9\plm2.9\km\\ 
\hline\rj\kak696.3\plm2.2\kak696.3\plm3.0\kak696.3\plm3.2\kak695.4\plm2.2\kak695.3\plm2.2\kak695.2\plm2.2\kak694.1\plm2.3\kak695.3\plm3.7\km\\ 
\hline
Minimization II & \multicolumn{8}{c}{\amud}  \\
\hline
$s_0$ values & LL & Pad\'e & Pol & GdR1 & GdR2 & GdR3 & GdR4 & GdR5 \\
\hline\rh\kak688.5\plm2.2\kak688.1\plm4.2\kak689.8\plm3.3\kak688.3\plm2.1\kak688.4\plm2.1\kak688.6\plm2.2\kak687.1\plm2.1\kak688.4\plm5.8\km\\ 
\hline\ri\kak689.5\plm2.2\kak689.1\plm4.2\kak690.8\plm3.3\kak689.3\plm2.1\kak689.4\plm2.1\kak689.6\plm2.2\kak688.1\plm2.1\kak689.4\plm5.7\km\\ 
\hline\rj\kak690.3\plm2.1\kak689.9\plm4.6\kak691.6\plm3.6\kak689.8\plm2.1\kak690.1\plm2.2\kak690.2\plm2.2\kak688.6\plm2.1\kak690.0\plm5.9\km\\ 
\hline
\end{tabular}\vglue2mm
\caption{Values of $a_\mu^{\text{HLO (I)}}$ for the parameterizations considered in this study and the three choices of $s_0$. Results using the two different minimization techniques for $K_1(s)$ are shown. Dataset I was used as input.}
\label{tab:2}
\end{table}

\newpage
\section{Evaluation of $a_\mu^{\text{HLO (II, III, IV)}}$}
The contour integral in $a_\mu^{\text{HLO (II)}}$ was calculated using the pQCD prediction of $\Pi_{had}(s)$. The QCD vector correlator $\Pi_\text{QCD}$, which is related to the vacuum polarization function via $\Pi_{had}(s)=-4\alpha\pi(\sum{q^{2}_i})\Pi_\text{QCD}(s)$, is known up to five loops in the massless approximation, i.e.\ to $O(\alpha_s^4)$~\cite{Baikov:2012zm}. The mass terms up to $O(\alpha_s^2(m^{2}/q^{2})^{30})$ are taken from~\cite{Maier:2011jd}, while the mass terms up to $O(\alpha_s^3(m^{2}/q^{2}))$ are taken from~\cite{Baikov:2009uw}. The strong coupling $\alpha_s(\mu^2)$ was determined from the PDG average for $\alpha_s(M_Z^2)$ using the CRunDec program~\cite{Herren:2017osy}. The uncertainty of the contour integral was estimated including contributions from the uncertainty of the input parameter $\alpha_s(M_Z^2)$, the variation of the renormalisation scale $\mu^2$ in the range from $s_0/2$ to $2s_0$, and the full value of the estimated duality violations~\cite{Pich:2016bdg,PhysRevD.103.034028}. These three terms have been added in quadrature to estimate the total uncertainty on the contour integral.
Note that for $s_0=(1.8\, {\rm GeV})^2$, the error is dominated by the estimate of the duality violations, which amount to 1\%, while the uncertainties from $\alpha_s$ and the scale variation are much smaller and of similar size. For $s_0=(2.5\,{\rm GeV})^2$, duality violations are already suppressed and contribute at the level of 0.1\%. They are negligible at $s_0=(12\,{\rm GeV})^2$, where the error is dominated by the uncertainty of $\alpha_s(M_Z^2)$.
To compute $a_\mu^{\text{HLO (III)}}$ and $a_\mu^{\text{HLO (IV)}}$, 
$e^+e^-$ data from Dataset I~\footnote{Similar values are obtained using Dataset II and III.} were used for $R(s)$ up to $s = \SI{10}{\,\giga\electronvolt}$, while pQCD was used above that value.
\newline Table~\ref{tab:3} shows the values of $a_\mu^{\text{HLO (II)}}$, $a_\mu^{\text{HLO (III)}}$ and $a_\mu^{\text{HLO (IV)}}$ for the three different $s_0$ values and the two minimization techniques for the approximated kernel.

\begin{table}[h]
\renewcommand{\arraystretch}{.9}
\begin{center}
\begin{tabular}{cccc}
\hline\hline
\multicolumn{4}{c}{Minimization I} \\ 
\hline
\sa\kak\amup\kak\amuq\kak\amur\km\\ 
\hline
\rh\kak2.94\plm0.04\kak0.43\plm0.01\kak2.95\plm0.05\km\\ 
\hline
\ri\kak1.84\plm0.01\kak-0.34\plm0.01\kak1.79\plm0.02\km\\ 
\hline
\rj\kak0.208\plm0.001\kak-1.695\plm0.035\kak0.079\plm0.001\km\\ 
\hline
\multicolumn{4}{c}{Minimization II} \\ 
\hline
\sa\kak\amup\kak\amuq\kak\amur\km\\ 
\hline
\rh\kak3.23\plm0.04\kak0.91\plm0.02\kak3.00\plm0.05\km\\ 
\hline
\ri\kak2.54\plm0.01\kak1.52\plm0.02\kak1.96\plm0.02\km\\ 
\hline
\rj\kak0.360\plm0.001\kak4.85\plm0.05\kak0.096\plm0.001\km\\ 
\hline
\hline
\end{tabular}\vglue2mm
\end{center}
\caption{Values of $a_\mu^{\text{HLO (II)}}$, $a_\mu^{\text{HLO (III)}}$, $a_\mu^{\text{HLO (IV)}}$ for the three choices of $s_0$ and the two different minimization techniques used to determine the approximated kernel $K_1(s)$. Dataset I was used as input.}
\label{tab:3}
\end{table}

\section{Results}
\begin{figure}[h!]
    \centering
    \includegraphics[width=0.49\textwidth,keepaspectratio]{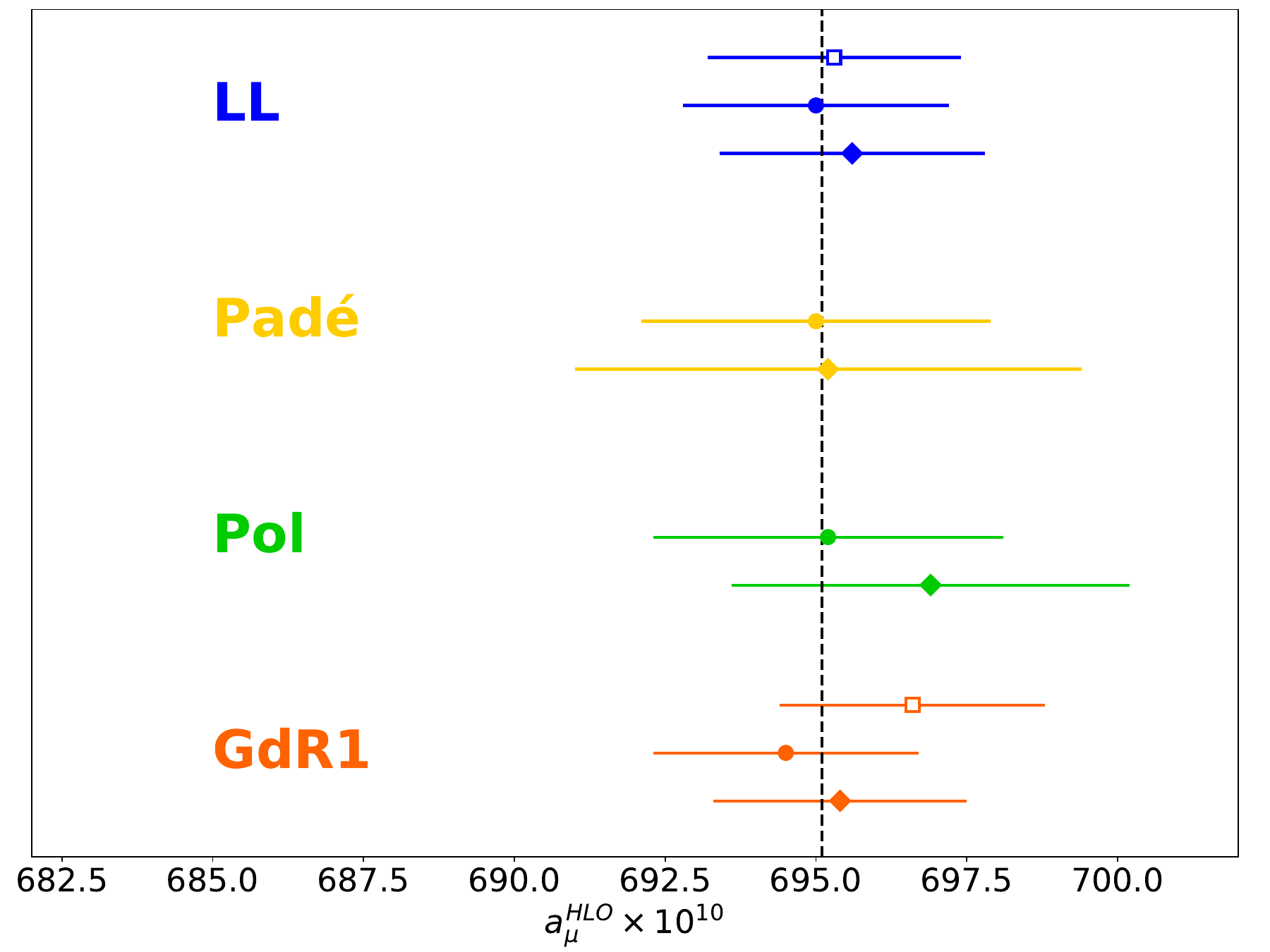}
    \includegraphics[width=0.49\textwidth,keepaspectratio]{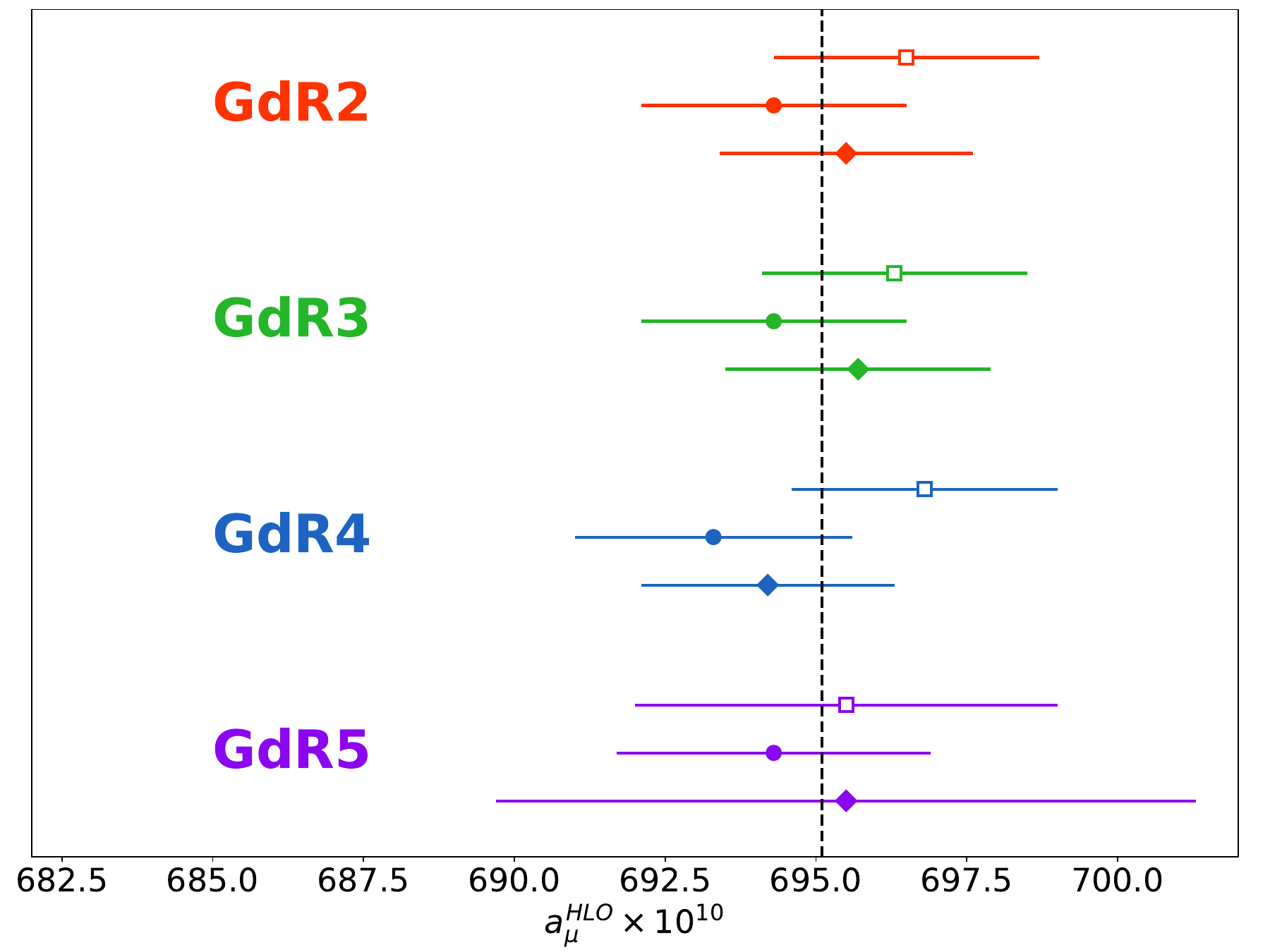}
    \caption{Values of \amuhlo \ for all parameterizations considered in this work, obtained from Dataset I with the different methods: MUonE integral from Eq.~\ref{eq:eqMUonEIntegral} (empty squares), our new method using Minimization 1 to get the approximated kernel (circles) and our new method using Minimization 2 (diamonds), using $s_0 = (\SI{1.8}{\,\giga\electronvolt})^2$. The results for the Pad\'e and Pol parameterizations computed using Eq.~\ref{eq:eqMUonEIntegral} are outside the plot range. The black dashed line represents the reference value obtained from Dataset I.}
    \label{fig:finalResults}
\end{figure}

Table~\ref{tab:4} shows the final results for \amuhlo \ obtained adding the four terms according to \mbox{Eq.~\ref{eq:eqAmuHLODer}}. They are compared to the values of \amuhlo \  obtained using the integral in Eq.~\ref{eq:eqMUonEIntegral}. Using this method, the results obtained from the Pad\'e and Polynomial parameterization are in strong disagreement with the reference input value. On the other hand, the same parameterizations can be safely adopted with our new method, since they allow to compute \amuhlo \ with no significant bias with respect to the reference value. For the sake of illustration, final results for $s_0 = (\SI{1.8}{\,\giga\electronvolt})^2$, Dataset I  and all the parameterizations are shown in Fig.~\ref{fig:finalResults}. Results obtained using Datasets II and III are very similar, and are reported in Tables~\ref{tab:5}, \ref{tab:6}.
%Tables~\ref{tab:5}, \ref{tab:6} and Figs.~\ref{fig:finalResults2}, \ref{fig:finalResults3} show the same results for \amuhlo \ obtained with our methods when Datasets II and III are used.

\section{Conclusions}
We have shown a method to compute \amuhlo\ with MUonE data using a derivative approach based on Refs.~\cite{Dominguez12,Dominguez17}. This method relies on the analyticity properties of the hadronic vacuum polarization function $\Pi_{had}(t)$ which allow to express  \dalphahad\ as a polynomial by a Taylor expansion at $t\sim0$.
The results obtained for different parameterizations used to fit  \dalphahad\ and different input datasets show that this method is competitive with the traditional one based on the integral of \dalphahad \ in the whole momentum transfer range. Moreover, the proposed method avoids the difficulties with the functional form of the parameterization used to extrapolate \dalphahad \ behaviour outside the MUonE range. We expect that by a convenient choice of the approximated kernel function this method can be applied to extract the NLO and NNLO hadronic vacuum polarization contributions to the muon \gmtwo\ in the space-like region.

\section*{Acknowledgements}
We thank Stefano Laporta, Eduardo de Rafael and David Greynat for useful discussions, and Massimo Passera for useful suggestions and contributions in the early phase of the project. This work was supported by the Leverhulme Trust, LIP-2021-01. TT is supported by the STFC Consolidated Grant ST/T000988/1.

%%%TABLE 4
\begin{table}[h]
\renewcommand{\arraystretch}{.9}
\begin{tabular}{ccccccccc}
\hline\hline
\multicolumn{9}{c}{\amuip}  \\
\hline
Minimization I & \multicolumn{8}{c}{\amude}  \\
\hline
$s_0$ values & LL & Pad\'e & Pol & GdR1 & GdR2 & GdR3 & GdR4 & GdR5 \\
\hline
\rh\kak 695.0\plm2.2\kak695.0\plm2.9\kak695.2\plm2.9\kak694.5\plm2.2\kak694.3\plm2.2\kak694.3\plm2.2\kak693.3\plm2.3\kak694.3\plm2.6\km\\ 
\hline
\ri\kak 695.0\plm2.2\kak694.9\plm3.0\kak695.1\plm3.0\kak694.3\plm2.2\kak694.1\plm2.2\kak694.1\plm2.2\kak693.1\plm2.3\kak694.2\plm2.9\km\\ 
\hline
\rj\kak 694.9\plm2.2\kak694.9\plm3.0\kak694.9\plm3.2\kak694.0\plm2.2\kak693.9\plm2.2\kak693.8\plm2.2\kak692.7\plm2.3\kak693.9\plm3.7\km\\ 
\hline
Minimization II & \multicolumn{8}{c}{\amude}  \\
\hline
$s_0$ values & LL & Pad\'e & Pol & GdR1 & GdR2 & GdR3 & GdR4 & GdR5 \\
\hline
\rh\kak 695.6\plm2.2\kak695.2\plm4.2\kak696.9\plm3.3\kak695.4\plm2.1\kak695.5\plm2.1\kak695.7\plm2.2\kak694.2\plm2.1\kak695.5\plm5.8\km\\ 
\hline
\ri\kak 695.5\plm2.2\kak695.1\plm4.2\kak696.8\plm3.3\kak695.3\plm2.1\kak695.4\plm2.1\kak695.6\plm2.2\kak694.1\plm2.1\kak695.4\plm5.7\km\\ 
\hline
\rj\kak 695.6\plm2.1\kak695.2\plm4.6\kak696.9\plm3.6\kak695.1\plm2.1\kak695.4\plm2.2\kak695.5\plm2.2\kak693.9\plm2.1\kak695.3\plm5.9\km\\ 
\hline
\hline
\amui\kak695.3\plm2.1\kak702.6\plm12.0\kak834.3\plm63.2\kak696.6\plm2.2\kak696.5\plm2.2\kak696.3\plm2.2\kak696.8\plm2.2\kak696.5\plm3.5\km\\ 
\hline
\hline
\end{tabular}\vglue2mm
\caption{Final results of \amuhlo \ calculated using the proposed method ($a_\mu^{\text{HLO (Der)}}$), for all parameterizations considered in this study, with the three choices of $s_0$ and the two minimization techniques used to determine the approximated kernel $K_1(s)$. $a_\mu^{\text{HLO (Int)}}$ is the result of the integral in Eq.~\ref{eq:eqMUonEIntegral}, when a specific parameterization for \dalphahad\ is used as input.  $a_\mu^{\text{HLO, INPUT}}$ is the reference value of \amuhlo \ obtained by integrating \dalphahad\ from Dataset I using Eq.~\ref{eq:eqMUonEIntegral}.}
\label{tab:4}
\end{table}

%%%TABLE 5 - FJ
\begin{table}[h!]
\renewcommand{\arraystretch}{.9}
\begin{tabular}{ccccccccc}
\hline\hline
\multicolumn{9}{c}{\amuipfj}  \\
\hline
Minimization I & \multicolumn{8}{c}{\amude}  \\
\hline
$s_0$ values & LL & Pad\'e & Pol & GdR1 & GdR2 & GdR3 & GdR4 & GdR5 \\
\hline
\rh\kak 690.3\plm2.2\kak690.2\plm2.9\kak690.5\plm2.9\kak689.7\plm2.2\kak689.6\plm2.2\kak689.6\plm2.2\kak688.6\plm2.3\kak689.6\plm2.6\km\\ 
\hline
\ri\kak 690.2\plm2.2\kak690.1\plm3.0\kak690.3\plm3.0\kak689.5\plm2.2\kak689.4\plm2.2\kak689.4\plm2.2\kak688.4\plm2.3\kak689.4\plm2.9\km\\ 
\hline
\rj\kak 690.1\plm2.2\kak690.0\plm3.0\kak690.0\plm3.2\kak689.1\plm2.2\kak689.1\plm2.2\kak689.1\plm2.2\kak688.1\plm2.3\kak689.1\plm3.7\km\\ 
\hline
Minimization II & \multicolumn{8}{c}{\amude}  \\
\hline
$s_0$ values & LL & Pad\'e & Pol & GdR1 & GdR2 & GdR3 & GdR4 & GdR5 \\
\hline
\rh\kak 690.8\plm2.2\kak690.4\plm4.1\kak692.1\plm3.3\kak690.7\plm2.1\kak690.7\plm2.1\kak690.8\plm2.2\kak689.4\plm2.1\kak690.7\plm5.9\km\\ 
\hline
\ri\kak 690.8\plm2.2\kak690.4\plm4.2\kak692.0\plm3.3\kak690.5\plm2.1\kak690.6\plm2.1\kak690.7\plm2.2\kak689.3\plm2.1\kak690.6\plm5.7\km\\ 
\hline
\rj\kak 690.8\plm2.1\kak690.4\plm4.6\kak692.1\plm3.6\kak690.4\plm2.1\kak690.5\plm2.2\kak690.6\plm2.2\kak689.1\plm2.1\kak690.5\plm5.9\km\\ 
\hline
\hline
\amui\kak690.5\plm2.1\kak695.8\plm8.8\kak724.2\plm31.2\kak691.7\plm2.2\kak691.7\plm2.2\kak691.6\plm2.2\kak692.0\plm2.2\kak691.7\plm3.5\km\\ 
\hline
\hline
\end{tabular}\vglue2mm
\caption{Same as Table~\ref{tab:4}, but using Dataset II as input.}
\label{tab:5}
\end{table}

%%%TABLE 6 - KNT
\begin{table}[h!]
\renewcommand{\arraystretch}{.9}
\begin{tabular}{ccccccccc}
\hline\hline
\multicolumn{9}{c}{\amuipknt}  \\
\hline
Minimization I & \multicolumn{8}{c}{\amude}  \\
\hline
$s_0$ values & LL & Pad\'e & Pol & GdR1 & GdR2 & GdR3 & GdR4 & GdR5 \\
\hline
\rh\kak 693.0\plm2.2\kak693.0\plm2.9\kak693.2\plm2.9\kak692.5\plm2.2\kak692.2\plm2.2\kak692.2\plm2.2\kak691.2\plm2.3\kak692.2\plm2.6\km\\ 
\hline
\ri\kak 692.9\plm2.2\kak693.0\plm3.0\kak693.1\plm3.0\kak692.3\plm2.2\kak692.1\plm2.2\kak692.1\plm2.2\kak691.0\plm2.3\kak692.1\plm2.9\km\\ 
\hline
\rj\kak 692.9\plm2.2\kak692.9\plm3.0\kak692.9\plm3.2\kak692.0\plm2.2\kak691.8\plm2.2\kak691.8\plm2.2\kak690.7\plm2.2\kak691.8\plm3.7\km\\ 
\hline
Minimization II & \multicolumn{8}{c}{\amude}  \\
\hline
$s_0$ values & LL & Pad\'e & Pol & GdR1 & GdR2 & GdR3 & GdR4 & GdR5 \\
\hline
\rh\kak 693.5\plm2.2\kak693.3\plm4.2\kak694.9\plm3.3\kak693.4\plm2.1\kak693.5\plm2.1\kak693.7\plm2.2\kak692.2\plm2.1\kak693.7\plm5.8\km\\ 
\hline
\ri\kak 693.5\plm2.2\kak693.3\plm4.2\kak694.8\plm3.3\kak693.3\plm2.1\kak693.4\plm2.1\kak693.6\plm2.2\kak692.1\plm2.1\kak693.6\plm5.7\km\\ 
\hline
\rj\kak 693.6\plm2.1\kak693.4\plm4.6\kak695.0\plm3.6\kak693.1\plm2.1\kak693.4\plm2.1\kak693.6\plm2.2\kak691.9\plm2.1\kak693.7\plm5.9\km\\ 
\hline
\hline
\amui\kak693.4\plm2.1\kak698.6\plm10.1\kak730.7\plm32.9\kak694.5\plm2.2\kak694.4\plm2.2\kak694.3\plm2.2\kak694.7\plm2.2\kak694.3\plm3.5\km\\ 
\hline
\hline
\end{tabular}\vglue2mm
\caption{Same as Table~\ref{tab:4}, but using Dataset III as input.}
\label{tab:6}
\end{table}

\FloatBarrier

\bibliography{biblio}

\end{document}